# Neutron-proton pairing in the unstable N=Z nuclei of the *f*-shell through two-nucleon transfer reactions

*M. Assié*[1*], *H. Jacob*[1], *Y. Blumenfeld*[1], *V. Girard-Alcindor*[1]

[1]Université Paris-Saclay, CNRS/IN2P3, IJCLab, 91405 Orsay, France.

**Abstract.** Pair transfer is a unique tool to study pairing correlations in nuclei. Neutron-proton pairing is investigated in the N=Z nuclei of the *f*-shell, through the reaction (p,$^3$He) in inverse kinematics, that allows to populate at the same time the lowest J=0$^+$, T=1 (isovector pairing) state and J=1$^+$, T=0 (isoscalar pairing) state. Radioactive beams of $^{56}$Ni and $^{52}$Fe produced by fragmentation at the GANIL/LISE facility combined with particle and gamma-ray detection make it possible to carry out this study from $^{48}$Cr (mid-shell nucleus) to $^{56}$Ni (doubly-magic nucleus). The cross-sections were extracted and compared with second-order distorted-wave born approximation (DWBA) calculations performed with neutron-proton amplitudes obtained from shell model calculations with GXPF1 interaction. Very low cross-sections for the J=1$^+$,T=0 state (isoscalar channel) were observed. The cross-section for $^{56}$Ni is one of order of magnitude lower than for $^{40}$Ca showing a strong reduction of the isoscalar channel in the *f*-shell as compared to the *sd*-shell. On the other hand, the increase of the cross-section towards the middle of the shell for the isovector channel points towards a possible superfluid phase.

## 1 General introduction

Pairing correlations are at the heart of superconductivity and superfluidity in strongly interacting quantum many-body systems. These phenomena span very different size scales, from color-superconducting quark matter [1] to neutron stars [2], and very different energies, from below meV in superconductors to MeV in nuclei and 100 MeV at the quark scale. Ultracold fermionic gases have brought deep insight into pairing as they offer the unique possibility to tune the strength of the pairing force via Feshbach resonances [3]. The theoretical description of pairing in all these quantum many-body systems is rooted in the microscopic theory of superconductivity developed by Bardeen, Cooper and Schrieffer (BCS) [4] with building blocks made of strongly correlated electron pairs, the Cooper pairs. Different attractive interactions underlie the Cooper pair mechanism: the electron-phonon interaction in conventional superconductors, the nucleon-nucleon interaction for nuclear superfluidity and the strong interaction in the case of color superconductivity. It is interesting to point out that for atomic nuclei, about half of the observed pairing gap is due to the coupling of particles with surface vibration phonons, in close analogy with the case of superconductors [5, 6]. Other effects may come into play to complete the description of quantum many-body systems, like quantum vorticity in superfluid neutron matter for neutron stars or complex dynamics of finite systems for nuclei.

### 1.1 Neutron-proton pairing

New types of superfluid fermionic states arise in multi-component systems with cross-species pairing. The most striking example arises for atomic nuclei where the neutron and proton fluids coexist. For most of the known nuclei, the superfluid states consist of spin-singlet (T=1) neutron and/or proton pairs (nn or pp pairs). In N=Z nuclei, the large overlap between neutron and proton wave functions favours another type of Cooper pairs made of neutron proton pairs (np pairs) of two different types: either isovector pairs (T=1) as for the nn and pp pairs or isoscalar (T=0) pairs, "deuteron-like". These latter are a very unique manifestation in nature of cross-species pairing. While the strength of the isoscalar pairing is strong enough to bind the deuteron, the question of the existence of a correlated state in analogy with the pair phase of superconductors remains opened [7] despite many experimental efforts and various observables like the binding energies, the rotational spectra and the two-nucleon transfer amplitudes.

### 1.2 Method: Two-nucleon transfer

In nuclei, single pair transfer has evidenced superfluid behaviour in Sn isotopes through two-neutron pick-up reactions (p,t) and two-neutron stripping reactions (t,p). In a superfluid nucleus, the typical number of Cooper pairs involved is of the order of 3 to 10, depending on the degeneracy of the specific orbitals involved, high-j orbitals being more favorable. Within a major shell, the enhancement of the pair transfer amplitude at mid-shell is driven by the ratio of the pairing gap to the strength of the pairing force, which reflects the number of Cooper pairs contributing coherently [8]. The enhancement of the cross-section to the lowest $0^+$ states ("pairing state") with respect to the second $0^+$ reached a factor of 25 in the Sn isotopes. Near the closed shells, pair vibrations sign the onset of the superfluid phase and the pair transfer cross-section is governed by the number of phonons involved [5,9]. It has been shown that atomic nuclei are very close of the transition point between the normal- and superfluid-phases [10].

---

[*] Corresponding author: marlene.assie@ijclab.in2p3.fr

The same arguments hold for np pair transfer [11-13]. In this case, two components enter into play: the deuteron-like (spin-triplet pair) and the spin-singlet pair transfer. Their interplay is probed starting from an even-even self-conjugate nucleus ($J=0^+$,$T=0$) and populating respectively the $J=1^+$,$T=0$ and the $J=0^+$,$T=1$ states in the residual odd-odd nucleus. The observable of interest is the ratio of the cross-sections: $\sigma(0^+)/\sigma(1^+)$.

The two-nucleon transfer reactions (p,$^3$He) and ($^3$He,p) are particularly well suited as their selection rules allow to populate both states at the same time. The np pair stripping ($^3$He,p) and np pair pick-up (p,$^3$He) transfers in direct kinematics in the stable N=Z nuclei of the *sd*-shell have been reviewed in [7] and recently remeasured consistently [14] (see Figure 1). Spin-triplet pairing remains elusive in the *sd*-shell due to the limited possible occupation and calls for further measurement in heavier nuclei. The comparison between np- ($^3$He,p) and two-neutron transfer (t,p) in the same shell is consistent with a strong dominance of the T=1 pairing. Tackling higher *j* orbitals and/or higher shells where N=Z nuclei are radioactive is very challenging given the low cross-sections and the radioactive beam intensities available. In this work, we investigate np pair transfer reactions through $^{56}$Ni(p,$^3$He)$^{54}$Co, $^{52}$Fe(p,$^3$He)$^{50}$Mn [15] and $^{48}$Cr(p,$^3$He)$^{46}$V to investigate spin-singlet and spin-triplet np pairing interplay in the *fp*-shell for the doubly-magic closed shell nucleus $^{56}$Ni, the open-shell nucleus, $^{52}$Fe and the mid-shell nucleus (and deformed) $^{48}$Cr.

## 2 Experimental set-ups and techniques

The three measurements have been performed in inverse kinematics at GANIL (France) within two experimental campaigns ($^{56}$Ni and $^{52}$Fe beams, on one side and $^{48}$Cr beams on the other side). The beams were produced by fragmentation of $^{58}$Ni and $^{50}$Cr and selection of the fragments of interest with the LISE spectrometer at 30 MeV/u. They were impinging of a thick CH$_2$ target (7 mg/cm$^2$ and 5 mg/cm$^2$ respectively). The light ejectile produced by the trnasfer reaction ($^3$He) is detected with the MUST2 array. Given that the projectile-like fragment are odd-odd nuclei, the density of states is such that particle only measurement does not allow to disentangle the populated states. This is why the MUST2 array was combined with EXOGAM clovers: 4 clovers in close configuration (8 cm from target) and 12 clovers in standard configuration (14 cm from target) respectively. The detection and identification in Z of the projectile-like nucleus at zero degree was added for the second campaign. The triple coincidence measurement ($^3$He, γ and projectile-like) improves greatly the background contribution. The set-up is supplemented by two beam tracking devices, CATS for beam counting and reconstruction of the vertex of the reaction on the target. More details on the experimental set-ups can be found in ref.[15] for the first campaign and in ref.[16] for the second campaign (MUGAST).

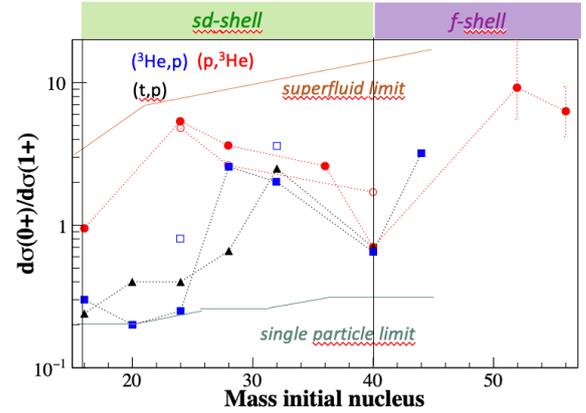

*Figure 1 : Systematics of the measured ratio of the cross-sections to the $0^+$,T=1 state over the cross-section to the $1^+$,T=0 state starting from an even-even N=Z nucleus as a function of the mass of the initial nucleus. The results for (3He,p) (np pair adding) (blue symbols) and (p,3He) (np pair removal) (red symbols) are displayed together with the ratio of the cross-sections to the first and second $0^+$ states for (t,p) (2n adding) reactions. The open symbols are coming from the consistent remeasurement for the stable isotopes of ref.[14].*

## 3 Extraction of the cross-sections to the lowest $0^+$ and $1^+$ states

### 3.1 $^{56}$Ni(p,$^3$He)$^{54}$Co and $^{52}$Fe(p,$^3$He)$^{50}$Mn

The excitation energy spectra obtained for $^{56}$Ni(p,$^3$He)$^{54}$Co and $^{52}$Fe(p,$^3$He)$^{50}$Mn, and shown in Figure 2, were extracted from the measurement of the $^3$He position and energy and using the two-body kinematics. The excitation energy resolution is of about 1 MeV mainly due to the contribution of the target thickness. The ground state ($J=0^+$, T=1 in both cases) is merged with the contribution of the excited states. The $J=1^+$, T=0 state is not identified by particle only, as expected. By applying a gate on the excitation energy centred on the energy of the $1^+$ state $\pm 2\sigma$, the gamma spectrum appears rather clean although with limited statistics. There is a clear contribution of the γ-ray associated to the $1^+$ state. Part of this is not related to the direct population of the $1^+$ state by the transfer reaction as most of the excited states above decay onto this state. By subtracting the top-feeding contribution and taking into account the efficiencies, the cross-section to the $1^+$ state can be deduced: for $^{54}Co$, $\sigma(1^+)=17\pm7$ μb and for $^{50}Mn$, $\sigma(1^+)=16^{+29}_{-16}$ μb. For the latter, the statistics is very low and the cross-section is compatible with 0.

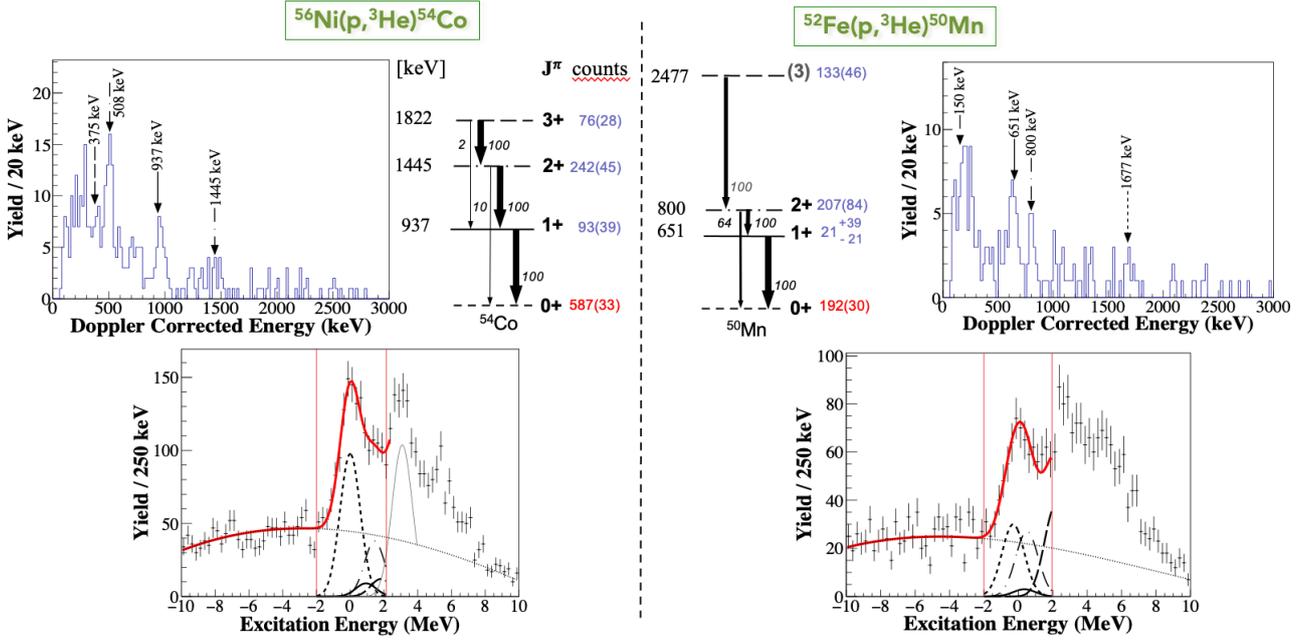

*Figure 2: Adapted from [15]. (Bottom) Excitation energy spectra obtained for both $^{56}Ni(p,^3He)^{54}Co$ (left) and $^{52}Fe(p,^3He)^{50}Mn$ (right) measurements. The black dotted line is the background contribution estimated from a $^{12}C$ run, the red line is the fit of the excitation energy spectrum with the components shown for each state. The only free parameter of the fit is the amplitude of the ground state peak. (Top) Associated gamma spectrum obtained by applying a gate on the excitation energy around the $1^+$ state $\pm 2\sigma$ (gate shown in red) and level scheme of the residual nuclei.*

As a second step, the cross-section to the $0^+$ ground state can then be determined by a fit of the excitation energy spectrum taking into account all the states identified by their γ-ray. Given that the energies of the states are known and the width of the peaks are known from simulations, the only free parameter is the amplitude of the ground state. The cross-sections for the $0^+$ ground state are respectively $\sigma(0^+) = 109 \pm 5$ μb for $^{54}Co$ and $\sigma(0^+) = 145 \pm 12$ μb for $^{50}Mn$.

### 3.2 $^{48}Cr(p,^3He)^{46}V$

The same principles have been used for the analysis of the third measurement with $^{48}Cr$ beam. The main difference lies in the use of the zero-degree detection (ZDD) of LISE [16] that allows to identify in Z the projectile-like nuclei. The ZDD is composed of a drift chamber (not used here), a set of 5 ionization chambers and a plastic covering from 0 to 2.5 degrees LAB. The Z identification can be performed by time-of-flight between the CATS beam tracking devices and the plastic of the ZDD versus the energy deposit in the five ionization chambers (see Figure 3). This supplemental information reduces the background contribution dramatically as it can be seen on Figure 4 where the kinematical lines are displayed and very few counts lie below the ground-state line. The ground state contribution is well separated from the contribution of the excited states. The second thick line includes several contributions including the $1^+$ state. To determine its cross-section, the triple coincidence with the γ-ray will be exploited.

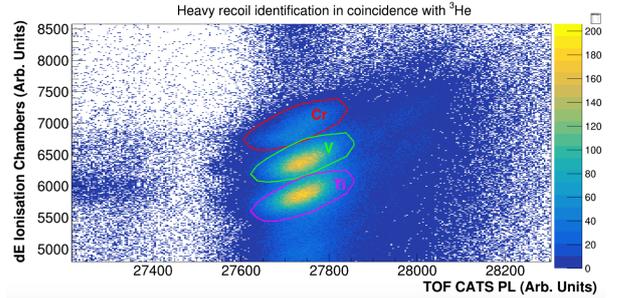

*Figure 3 : Identification of the projectile-like residual nucleus in the ZDD by using the time-of-flight between the CATS beam tracking devices and the plastic of the ZDD and the summed energy deposit in the five ionization chambers. The spectrum is obtained in coincidence with an $^3He$ detected in MUST2.*

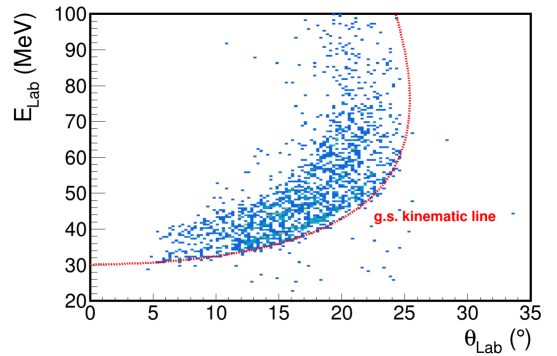

*Figure 4 : Preliminary kinematical lines (energy of the detected $^3He$ versus its angle in the laboratory frame) obtained for the reaction $^{48}Cr(p,^3He)^{46}V$.*

## 4 Results and discussion

The ratios of the cross-sections for $^{56}Ni(p,^3He)^{54}Co$ and $^{52}Fe(p,^3He)^{50}Mn$ are displayed in Figure 1 including statistical and systematic error bars. The comparison with the *sd*-shell reveals that the ratios are higher in the *f*-shell and particularly does not go down to 1, as expected, in the doubly-magic nucleus $^{56}$Ni. It reflects that fact the $1^+$ state is weakly populated in the *f*-shell.

It has also been noticed in the measurement of $^{44}$Ti($^3$He,p)$^{46}$V [7] but not in the two-nucleon transfer measurements performed in the sd-shell. The higher value is probably due to the weakening of the T=0 np pairing in the *fp*-shell due to the spin-orbit effect [17-20].

The ratios were also compared to DWBA calculations (see details in ref.[15]) including two-nucleon amplitudes (TNA) for two cases : the single particle case (where no extra-correlation between the nucleons is added) and the "pairing" case where the TNA where extracted from shell model calculations performed with the GXPF1 interaction. The results are consistent with the theoretical estimates in the pairing case, although the large experimental error bars may not exclude a compatibility with the single particle case (particularly in the case of $^{56}$Ni). The comparison with $^{40}$Ca(p,$^3$He) from ref.[14] shows that when the two nucleons are picked up from the *sd*-shell, the isoscalar pairing dominates the isovector pairing (the experimental point, with error bars smaller than the dot, is below the single particle prediction) whereas when picking them up from the *fp*-shell, the situation is opposite. It confirms the hindering of the isoscalar pairing in the *fp*-shell with respect to the *sd*-shell.

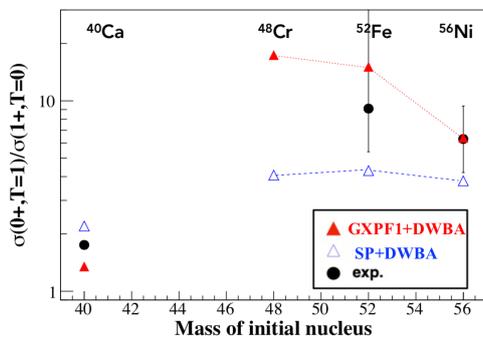

*Figure 5 : Systematics of the measured ratio of the cross-sections to the $0^+$,T=1 state over the cross-section to the $1^+$,T=0 state of the f-shell. The experimental results are shown with black dots together with the theoretical predictions in the single particle assumption (blee triangles) and in the pairing assumption (red triangles) (taken from ref. [15]).*

The experimental results for the $^{48}$Cr(p,$^3$He) reaction are not yet ready for publication but the theory predicts an increase of the ratio at the middle of the shell. However, this effect may be strongly modified by the deformation of $^{48}$Cr [21].

## 5 Conclusion

In this paper, we report on recent measurements along the *f*-shell to investigate the role of both the isoscalar and the isovector pairing. The second half of the *f*-shell has been covered via the two-nucleon transfer reaction (p,$^3$He) performed in consistent way. Isovector neutron-proton pairing plays an important role whereas the isoscalar neutron-proton pairing remains very elusive. Either a much larger shell occupancy is needed for strong correlations to develop or its strength is too fragmented among the surrounding quasi-particle state backgrounds to be observed. Further exploration towards heavier N=Z nuclei in the *g*-shell will have to wait for the new generation of radioactive beam facilities.